\documentclass[ms]{copernicus}
\def\simle{\lower 2pt \hbox {$\buildrel < \over {\scriptstyle \sim }$}}
\def\simge{\lower 2pt \hbox {$\buildrel > \over {\scriptstyle \sim }$}}

\def\ep{E_{\rm p}}

\def\enu{E_{\nu}}

\usepackage[dvips]{epsfig}

\begin{document}

%\linenumbers

\title{Neutrinos from Colliding Wind Binaries: 
Future Prospects for PINGU and ORCA}

\author[1]{Julia Becker Tjus}
%\author[]{NAME}
%\author[]{NAME}

\affil[1]{Theoretische Physik IV: Plasma-Astroteilchenphysik, Fakult\"at f\"ur Physik \& Astronomie, Ruhr-Universit\"at Bochum 44780 Bochum, Germany}
%\affil[]{ADDRESS}

%% The [] brackets identify the author to the corresponding affiliation, 1, 2, 3, etc. should be inserted.

\runningtitle{Neutrinos from Colliding Wind Binaries}

\runningauthor{Julia Tjus}

\correspondence{Julia Becker Tjus\\ (julia.tjus@rub.de)}

\received{}
\pubdiscuss{} %% only important for two-stage journals
\revised{}
\accepted{}
\published{}

%% These dates will be inserted by the Publication Production Office during the typesetting process.

\firstpage{1}

\maketitle  %% Please note that for the copernicus2.cls this command needs to be inserted after \abstract{TEXT}

\begin{abstract}
Massive stars play an important role in explaining the cosmic ray spectrum below the knee, possibly even up to the ankle, i.e.\ up to energies of $10^{15}$~eV or $10^{18.5}$~eV, respectively. 
In particular, Supernova Remnants are discussed as  one of the main candidates to explain the cosmic ray spectrum. 
Even before their violent deaths, during the stars' regular life times, cosmic rays can be accelerated in wind environments. 
High-energy gamma-ray measurements indicate hadronic acceleration binary systems, leading to both periodic gamma-ray emission from binaries like LSI+60~303 and continuous emission from colliding wind environments like Eta Carinae. 
The detection of neutrinos and photons from hadronic interactions are one of the most promising methods to identify particle acceleration sites. 
In this paper,  future prospects to detect neutrinos from colliding wind environments in massive stars are investigated. In particular, the seven most promising candidates for emission from colliding wind binaries are investigated to provide an estimate of the signal strength.
The expected signal of a single source is about a factor of $5-10$ below the current IceCube sensitivity and it is therefore not accessible at the moment. 
What is discussed in addition is future the possibility to measure low-energy neutrino sources with detectors like PINGU and ORCA: the minimum of the atmospheric neutrino flux at around $25$~GeV from neutrino oscillations provides an opportunity to reduce the background and increase the significance to searches for GeV/TeV neutrino sources. This paper presents the first idea, detailed studies including the detector's effective areas will be necessary in the future to test the feasibility of such an approach.
\end{abstract}

\introduction  %% \introduction[modified heading if necessary]
While the sources of high-energy cosmic rays are still not directly identified, the search for neutrinos and photons from cosmic ray interactions in the vicinity of the acceleration environments has made great progress within the past couple of years: 
\begin{enumerate}
\item  The {\it Fermi} satellite was able to measure gamma-ray emission from two SNRs, i.e.\ W44 and IC443, in accordance with hadronic models \citep{fermi_snrs2013}: {\it Fermi} detects gamma-rays from $\sim 100$~MeV up to GeV energies and is therefore sensitive to the low-energy cutoff from the pion induced gamma-ray spectrum at $\sim 200$~MeV. Such a cutoff was reported in \cite{fermi_snrs2013} for the two SNRs W44 and IC443, providing a first piece of evidence for the hadronic nature of the signal. The detection of these two SNRs, however, does not provide information about the entire observed cosmic ray budget. The spectra of these two SNRs in particular are much too steep toward TeV energies to be able to explain the observed cosmic ray spectrum with a spectral behaviour of $E^{-2.7}$, see e.g.\ \cite{matthias_snr2013}.
\item  A first signal of high-energy neutrinos (TeV-PeV energies) was recently announced by the IceCube collaboration \citep{icecube_evidence2013}. The signal shows an astrophysical flux up to PeV energies, which either represents an $E^{-2}$ flux which cuts off toward higher energies {\it or}  a somewhat steeper flux ($\sim E^{-2.2}$) persisting to higher energies \citep{icecube_evidence2013}. At this point, this first signal is fully consistent with an isotropic flux, so no individual sources have been identified yet. The possible cutoff at PeV energies makes the interpretation as Galactic cosmic ray sources an interesting option. This would mean that a clustering of the events within the Galactic plane will have to be observed in the future. A purely isotropic distribution would, on the other hand, rather point to an extragalactic flux.
\end{enumerate}

Cosmic ray energy spectra from stellar environments are believed to be of too low energy in order to explain the total flux of cosmic rays up to the knee: the observed cosmic ray energy spectrum is believed to be produced by one dominant source class up to the first, most prominent change in the spectral behaviour of the cosmic ray spectrum, the so-called knee at $10^{15}$~eV. Stellar environments cannot provide such extreme energies, it is rather expected to have particle acceleration up to $100$~TeV energies at maximum, something that is obvious from the Hillas criterion, see e.g.\ \cite{becker2008} for a summary. Thus, stellar sources must be sub-dominant with respect to the entire cosmic ray budget.
 Cosmic rays from stars could, however, contribute to the neutrino and gamma-ray flux by interacting with the stellar wind environment. A summary of different neutrino emission scenarios from massive stars is given in \cite{romero2010}, gamma-ray emission from Wolf-Rayet binaries is discussed in detail in \cite{benaglia_romero2003}. In particular, colliding winds in massive binary systems provide a good source for both cosmic ray acceleration, due to the shock front formed by the colliding winds, as well as particle interaction, due to the dense wind environments \citep{eichler_usov1993}.  The prominent case of $\eta$ Carinae was detected at gamma-ray energies by the {\it Fermi} satellite \citep{fermi_etacar2010}, showing both a steep component possibly representing inverse Compton photons \citep{reimer2006} and a flat component persisting towards higher energies \citep{farnier2011}. The latter is a strong indication for hadronic acceleration up to GeV energies. Gamma-rays produced in proton-proton interactions are always accompanied by neutrinos. For the case of Eta-Carinae, this is already discussed in detail in \cite{bednarek_pabich2011}. In this paper, the high-energy neutrino flux from colliding wind binary systems (CWBs) is calculated using most recent limits and detections with the Fermi satellite. The very low cosmic ray maximum energy of below $100$~TeV leads to neutrino maximum energies of around a few TeV ($E_{\nu}\sim E_{p}/20$). It is thus clear from the beginning that a high-energy neutrino telescope like IceCube with a lower energy threshold of $100$~GeV and maximum performance at above $10$~TeV neutrino energy will have a difficult time to detect such signatures. Low-energy extensions like PINGU and ORCA, on the other hand, might provide a possibility of how to detect these sources. This idea will be discussed in detail in this paper.

%%%%%%%%%%%%%%%%%%%%%%%%%%%%%%%%%%%%%%%%%%%%%%%%%%%%%%%%%%%%%%%%%%%%%%%%%%%%%%%%%%%%%%%%%%%%%%%%%%%%%%%%%%
\section{Colliding wind binaries: neutral secondaries from cosmic ray interactions}
%%%%%%%%%%%%%%%%%%%%%%%%%%%%%%%%%%%%%%%%%%%%%%%%%%%%%%%%%%%%%%%%%%%%%%%%%%%%%%%%%%%%%%%%%%%%%%%%%%%%%%%%%%
When hadronic cosmic rays interact with a matter target locally at the cosmic ray acceleration site, neutral particles like gamma-rays and neutrinos are produced via pion- and kaon decay from the interaction \citep[e.g.]{becker2008}. In astrophysical environments, it is usually sufficient to include the contribution from pions and neglect the kaons as well as any contribution from charmed particles. The main reason is the low target density, which does not allow the pions to interact before they decay. Thus, the much more frequent process of pion decay usually dominates the neutrino spectrum. Situations of extreme environment can change that, as can be seen at the example of gamma-ray bursts, discussed in e.g.\ \cite{winter2012}. In general, proton-proton and proton-photon interactions produce neutrinos in astrophysical environments. Here, we focus on proton-proton interactions due to the high matter density in stellar wind environments which should dominate the total optical depth for hadronic interactions. An additional reason is the very high energy threshold for neutrino-production via proton-photon interactions, see e.g.\ \cite{becker2008} for a detailed discussion. With maximum energies around some TeV, it is not expected to have a significant contribution. 
%The threshold to produce a Delta-resonance via proton-photon interactions is $E_p>(m_{\Delta}^{2}-m_{p}^{2})/(4\,<E_{\gamma}>$, with $<E_

The neutrino spectrum from charged pion-decay is given as
\begin{equation}
\Phi_{\nu}(E_{\nu})=c\cdot n_H\cdot \int_{0}^{1} \sigma_{\rm pp}(\enu/x)\cdot J_{\rm p}(E_{\nu}/x)\cdot F_{\nu}(x,\enu/x)\cdot \frac{dx}{x}\,.
\end{equation}
Here, the integration is performed in $x=E_{\rm p}/\enu$. Functions are the total inclusive proton-proton cross-section $\sigma_{\rm pp}$, the number of protons per unit area and energy at the source $J_{\rm p}$ and the distribution of neutrino production from a single interaction of a proton at energy $\ep$, $F_{\nu}$. The factor $dx/x$ takes into account the relative contribution of protons in the interval $(d\ep,\,\ep+d\ep)$. An analytic approximation derived from numerical interaction models can be found in \cite{kelner2006} and will be used here.

Neutrinos are produced both directly from the charged pion decay, $\pi\rightarrow \mu\,\nu$, and from the muon decay, $\mu\rightarrow e\,\nu\,\nu$. Each of the three neutrino fluxes receives approximately the same amount of total energy from the pion. Thus, on total, $\sim 3/4$ of the energy is going into the neutrinos. Once produced, neutrinos propagate straight without further interaction, and the only thing which needs to be taken into account is their oscillation. A ratio of $(\nu_{e}:\nu_{\mu}:\nu_{\tau})=(1:1:1)$ is expected on average at such long distances, when starting with the pion-induced ratio of $(1:2:0)$ at the source.

The uncertainties of the calculation from astrophysics, in particular the unknown cosmic ray flux at the source, but also the target density, dominate the total uncertainties of the calculation. In this context, ingredients from particle physics (i.e.\ neutrino distribution function and proton-proton cross-section) are sufficiently well-known. In this paper, gamma-ray observations are used to constrain the astrophysical parameter space. 

The spectral behaviour of the neutrinos and photons follows the cosmic ray spectrum directly above a threshold energy of $\sim 200$~MeV and up to a maximum energy of about $1/20$ for neutrinos and $1/10$th for photons of the maximum cosmic ray energy. Diffusive shock acceleration indicates cosmic ray spectra at the source of $\sim E^{-2}$, see e.g.\ \cite{gaisser1990} for a summary. For simplicity, an $E^{-2}$ spectral behaviour is used here, with a cosmic ray maximum energy of 100~TeV.

A binary system is described via its mass loss rates $\dot{M}_i$ ($i=1,2$)
and wind velocities $V_i$. The colliding winds provide a total power of $L_{W}=1/2\cdot \left(\dot{M}_{1}\cdot V_{1}^2+\dot{M}_{2}\cdot V_{2}^2\right)$ and it is expected that parts of this power is going into the acceleration of cosmic rays.
Cosmic rays are therefore limited in their total luminosity $L_{cr}$ by the energy available from the total wind power $L_{W}$:
total wind power:
\begin{equation}
L_{cr}=\eta\cdot L_{W}=\frac{\eta}{2}\cdot \left(\dot{M}_{1}\cdot V_{1}^2+\dot{M}_{2}\cdot V_{2}^2\right)
\end{equation}
with $\eta<1$.
For the seven most interesting systems, among them $\eta$~Car, WR140 and WR147,
parameters are listed in Table \ref{wind_parameters:tab}.
\begin{table}
\centering{
\begin{tabular}{l||lllllll}
\hline
&$d$&$\dot{M}_{1}$&$V_{1}$&$\dot{M}_{2}$&$V_{2}$&$n_H$\\
&[kpc]&[$10^{-5}M_{\odot}/$yr]&[km/s]&[$10^{-7}M_{\odot}/$yr]&[km/s]&[$10^{37}$~erg/s]&\\\hline\hline
$\eta$ Car&2.3&$25 $&$500$&$100$&$3000$&$\sim 10^{9}$cm$^{-3}$\\\hline
WR11&$0.41$&$3$&$1550$&$1.78$&$2500$&$<0.07\cdot n_{H}^{\eta-car}$\\
WR137&$2.38$&$3.3$&$1900$&N/A (0)&N/A (0)&$<0.1\cdot n_{H}^{\eta-car}$\\
WR140&$1.85$&$4.3$&$2860$&$87$&3100&$<0.1\cdot n_{H}^{\eta-car}$\\
WR146&$1.25$&$4$&$2700$&$80$&$1600$&$<0.2\cdot n_{H}^{\eta-car}$\\
WR147&$0.78$&$2.4$&$950$&$4$&$800$&$<0.1\cdot n_{H}^{\eta-car}$\\\hline
Arches Cluster&$8.33$&$60$&$1500$&--&--&$\sim 10^{8}$~cm$^{-3}$\\\hline
\end{tabular}
\caption{Wind parameters. All values for $\eta$~Carinae are taken from \cite{reimer2006,farnier2011}, all parameters for Wolf-Rayet binaries, except for the densities, are taken from \cite{werner2013} and references therein. As a conservative limit, we use the maximum allowed distance for the sources (including errors), leading to the lowest possible neutrino flux. The remaining free parameter in the calculation is the target density, which is chosen in order to obey the gamma-ray limits from \cite{werner2013}. For the Arches cluster, we follow \cite{raga2001} in assuming a number of 60 CWBs with a mass loss rate of $\dot{M}_{\rm W}\sim 10^{-4}\,M_{\odot}/$yr and a wind velocity of $v_{\rm W}\sim 1500$~km/s each, and using the estimated distance of $8.33$~kpc. The density is taken to be $10^{8}$~cm$^{-3}$, which is a factor of $10$ above what is cited in \cite{figer2002} in order to account for the increased density in the shock region, as discussed in \cite{farnier2011}. \label{wind_parameters:tab}}
}
\end{table}

The total hadronic gamma and neutrino power scales with three physical input parameters:
\begin{enumerate}
\item {\bf Wind luminosity:} under the assumption that a fixed budget of the total wind velocity, $L_{\rm CR}=\eta\cdot L_{\rm W}$, the neutrino flux scales with this variable, $\phi_{\nu}\propto L_{\rm W}$.
\item {\bf Hydrogen density:} the fraction of protons interacting in the gas depends linearly on the target density assuming optical depths less than 1, which is a reasonable assumption. Therefore, neutrino production from a fixed cosmic ray flux depends on the hydrogen density $n_{\rm H}$, where all hydrogen species (H-I, H-II and H$_2$ are counted): $\phi_{\nu}\propto n_{H}$. It should be noted that the presence of helium in the wind would further increase the signal. The flux would scale up by a factor $(1+n_{\rm HE}/n_{\rm H})$, where $n_{\rm He}$ is the local helium density. As uncertainties on the fraction of helium at the source are rather large, this additional contribution is neglected at this point.
\item {\bf distance to the source:} the flux observed at earth then scales with one over the distance from the source squared: $\phi_{\nu}\propto 1/d_{L}^2$.
\end{enumerate}
Thus, on total, the neutrino flux scales with the variable
\begin{equation}
\chi_{\nu}=\frac{L_{W}\cdot n_H}{d_{L}^{2}}=10^{46}\frac{\rm erg}{{\rm s\,cm}^3\,{\rm kpc^{2}}}\cdot \left(\frac{L_{W}}{5\cdot 10^{37}\,{\rm erg/s}}\right)\cdot \left(\frac{n_{H}}{10^{9}\,{\rm cm}^{-3}}\right) \cdot \left(\frac{d_{l}}{2\,{\rm kpc}}\right)^{-2}
\end{equation}
The number of $10^{46}\frac{\rm erg}{{\rm s\,cm}^3\,{\rm kpc^{2}}}$ given in the above equation represents the value for $\eta$-Car with the input-parameters as indicated. In this case, the expected neutrino flux can be derived directly from the measurements of the high-energy gamma-ray component \citep{farnier2011}.  Wind parameters for WR11, WR137, WR140, WR146 and WR147 are taken from \cite{werner2013}, who also provide restrictive limits to the gamma-ray flux from measurements with the {\it Fermi} satellite. In this paper, these limits are used to normalise the photon and neutrinos fluxes, a procedure that leads to the prediction of an upper limit to the density in the source $n_H$: as all parameters are fixed in the calculation according to Table \ref{wind_parameters:tab}, the density needs to be reduced with respect to the density in $\eta$ Carinae to obey the limits.

\begin{figure}
\centering{
\epsfig{file=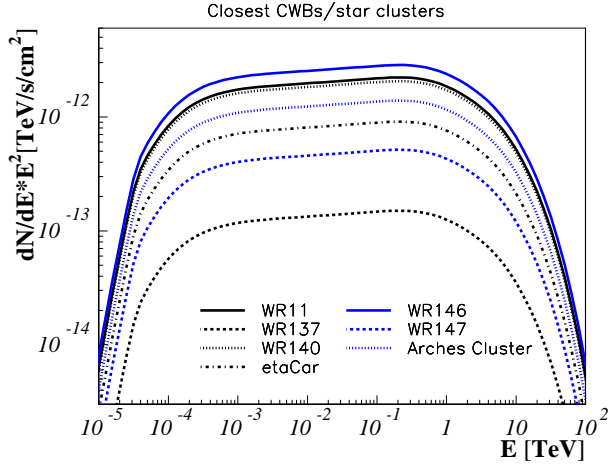,width=0.6\linewidth}
\caption{Muon neutrinos from Wolf-Rayet Binaries, Eta Carinae and the Arches Cluster. The fluxes of the Wolf-Rayet binaries are fixed to obey limits from gamma-ray observations.\label{eta_carinae_numu}}
}
\end{figure}

Figure \ref{eta_carinae_numu} shows the resulting neutrino fluxes. As described above, the upper limit from observations to the distance of the source is used in the calculations, resulting in a conservative estimate of the fluxes. Then, the density is estimated in an optimistic fashion, choosing a value close to the limit for gamma-ray emission given in \cite{reimer2006}. 

\conclusions[Conclusions and Outlook]

Figure \ref{eta_carinae_numu} shows that potential CWB neutrino sources range from flux levels of $E^2dN/dE\sim 10^{-13}-2\cdot 10^{-12}$~GeV/(cm$^{2}$~s~sr). For one year of operation, the sensitivity of the IceCube  between $0.1$~TeV and $10$~TeV is at a level of $\left.E^2dN/dE\right|_{\rm sens}\sim 10^{-13}-2\cdot 10^{-12}$~GeV/(cm$^{2}$~s~sr) for the first three years of operation, depending on the exact declination of the source \citep{icecube_ps2013}. IceCube is sensitive to northern hemisphere sources at these energies, i.e.\ to WR137, WR140, WR146 and  WR147. Sources in the southern hemisphere, i.e.\ WR11, eta Carinae and the Arches cluster, are interesting for the operating ANTARES telescope, but in particular for the KM3NeT observatory with its improved sensitivity.
From the first IceCube results \citep{icecube_ps2013}, it is clear that a detection of neutrinos from CWBs cannot be accommodated right now. Also, the detected, diffuse flux of astrophysical neutrinos cannot arise from these sources, as it persists up to PeV energies, so much higher than expected for CWBs, where the maximum neutrino energy would rather be in the GeV-TeV range, assuming that neutrinos receive approximately $1/20$th of the initial cosmic ray energy, see e.g.\ \cite{becker2008} and references therein.

The central question is now how the detection probability could be enhanced when concerning CWBs. Apart from considering the obvious analysis strategies like searching for individual point sources or stacking the brightest sources, it is most important to exploit the signal at low energies as efficiently as possible. Within IceCube, the DeepCore array enlarges the effective area at energies around $30$~GeV by a factor of larger than $2$ \citep{deep_core2012}. In the future, the low-energy IceCube extension PINGU will improve the sensitivity at GeV energies even further \citep{pingu_loi}. What could be exploited in this context is the probability {\it minimum} for $\nu_{\mu}\rightarrow \nu_{\mu}$ oscillations at $\sim 25$~GeV, i.e.\ $P(\nu_{\mu}\rightarrow\nu_{\mu}, E_{\nu}\sim 25$~GeV$)=0$, see \cite{pingu_loi} and references therein. In a purely theoretical view, the atmospheric muon neutrino flux is reduced to zero at these energies. This would, again theoretically, reduce the background to zero, such that a single signal event would become significant. Of course, limited energy resolution and other uncertainties connected to the detection of GeV neutrinos will make such a detection strategy much more complicated. One other important issue is the precise estimate of the atmospheric neutrino flux at GeV energies. The use of CORSIKA low-energy extensions (e.g.\ Fluka) make it possible to model the atmospheric flux with up-to-date fits of the measured cosmic ray flux at Earth and current cross sections. This work is ongoing and it will provide a tool in the future to reduce the uncertainties from the theoretical side to a minimum, but certainly not to zero.

While the above uncertainties make a detection of a GeV-signal from astrophysical sources difficult, it could still be feasible. This option should be considered for future detection arrays like PINGU, as it may be the only chance to see neutrinos from cosmic ray sources with maximum energies smaller than some TeV. One advantage could be that searches for these sources do not concern the entire sky, but can be limited to the region where the source/the sources are located, thus reducing the background to a minimum.

To summarise, as of {\it today}, the detection of neutrinos from CWBs with IceCube or ANTARES is rather unlikely due to a relatively low flux in combination with a low maximum energy. With the first detection of an astrophysical signal, high-energy neutrino astrophysics has just begun. Therefore, for the {\it future}, considering the potential of low-energy extensions like PINGU or ORCA, CWBs should be certainly be considered as interesting neutrino sources, having the potential to reveal both acceleration properties as well as information about the local hydrogen density.

\begin{acknowledgements}
I would like to thank John Black, Ralf Kissmann, Horst Fichtner and Klaus Scherer for valuable discussions on this topic before, during and after this workshop. In addition, thanks to the two anonymous referees for providing constructive and valuable comments. This work was partially supported by the DFG project BE3714/4-1, the ASPERA project {\it A Concerted R\&D Program for Low Energy Neutrino Detectors} and the research department of plasmas with complex interactions (Bochum).
\end{acknowledgements}

\end{document}